\documentclass[11pt,a4paper]{article}
\usepackage[utf8]{inputenc}
\usepackage{amsmath}
\usepackage{amsfonts}
\usepackage{amssymb}
\usepackage{enumerate}   
\usepackage{enumitem}   
\usepackage{fancyvrb}

\newcommand{\nL}{\mathcal{L}}
\newcommand{\ee}{\mathrm{e}}

\renewcommand{\AA}{\mathtt{A}}
\newcommand{\BB}{\mathtt{B}}
\newcommand{\CC}{\mathtt{C}}

\newcommand{\QQ}{\mathbb{Q}}

\begin{document}

\begin{center}
    {\LARGE User manual for {\tt bch}, a program for the fast computation of the\\[2mm]
    Baker--Campbell--Hausdorff and similar series}
\vskip 20pt
{\bf Harald Hofst\"atter}\\ 
{\small\it Reitschachersiedlung 4/6, 7100 Neusiedl am See, Austria}\\
{\tt hofi@harald-hofstaetter.at} 
\vskip 15pt
Program version 1.2, \today
\end{center}
\vskip 20pt

\centerline{\bf Abstract}
\noindent
This manual describes \verb|bch|, 
an efficient  program written in the C programming language for the fast computation
of the Baker--Campbell--Hausdorff (BCH) and similar Lie series.
The Lie series can be represented in the Lyndon basis, in the
classical Hall basis, or in the right-normed basis of 
E.S.~Chibrikov.  In the Lyndon basis,
which proves to be particularly efficient for this purpose,
the computation of 111\,013 coefficients for the BCH series up to terms of degree 20
takes less than half a second on an ordinary personal computer and requires negligible 11\,MB of memory.
Up to terms of degree 30, which is the maximum degree the program can handle, 
the computation of 74\,248\,451 coefficients takes 55 hours but still requires only a modest 5.5\,GB of  memory.

\vskip 30pt

\section{Introduction}
We consider the element 
\begin{equation}\label{eq:BCH_element}
H = \log(\ee^{\AA}\ee^{\BB})
= \sum_{k=1}^\infty\frac{(-1)^{k+1}}{k}\big(\ee^{\AA}\ee^{\BB}-1\big)^k
= \sum_{k=1}^\infty\frac{(-1)^{k+1}}{k}\bigg(\sum_{i+j>0}\frac{1}{i!j!}\AA^i\BB^j\bigg)^k 
\end{equation}
in the ring $\QQ\langle\langle\AA,\BB\rangle\rangle$  of formal power series 
in the non-commuting variables $\AA$ and $\BB$ with rational coefficients.
This element $H$ is commonly called the Baker--Campbell--Hausdorff (BCH) series. 
A classical result known as the
 Baker--Campbell--Hausdorff theorem (see, e.g.,~\cite{HHproof}) states that 
$H$
is a Lie series, which means that the 
homogeneous components $H_n$ of degree $n=1,2,\dots$ in
\begin{equation}
H=\sum_{n=1}^\infty H_n, 
\end{equation}
can be written as  linear combinations of 
$\AA$ and $\BB$ and (possibly nested) 
 commutator terms in $\AA$ and $\BB$, i.e., they
 are 
elements of the
free Lie algebra $\nL_\QQ(\AA,\BB)$ 
generated by $\AA$ and $\BB$.  

The program \verb|bch| computes the terms of the Lie series $H$ 
up to terms of a given maximal degree $N$, where $H$ (or more precisely, each
component $H_n$, $n=1,\dots,N$) is optionally represented in the Lyndon basis, 
the right-normed basis
of E.S.~Chibrikov \cite{Chibrikov},\footnote{The right-normed 
Chibrikov basis consists of right-normed 
(i.e., right-nested) commutators of the form $[a_1,[a_2,\dots[a_{n-1},a_{n}]\dots]]$
where $a_i\in\{\AA,\BB\}$. It seems that there are no
reports in the literature on computations of the BCH series in terms of 
linearly independent sets of such right-normed commutators
up to high degree. In \cite{ArnalCasasChiralt} such computations are performed
only up to degree $N=10$, and the approach taken there does not seem to be as
systematic, general and efficient as that of our
{\tt bch} program, which has computed
the BCH series in the right-normed Chibrikov basis up to degree $N=22$, 
see Section~\ref{Sect:performance}.}
or the classical Hall basis. Furthermore, the program can straightforwardly
be extended to compute the representation in an arbitrary generalized Hall basis.

In addition to $H=\log(\ee^\AA\ee^\BB)$ the program  \verb|bch| can  compute  
the Lie series   for arbitrary expressions of the form 
\begin{equation}\label{eq:expression}
  X = \log(\ee^{\Phi_s}\cdots\ee^{\Phi_1}),
  \end{equation}
   where the $\Phi_i$ are Lie polynomials 
with rational coefficients in two or more non-commuting variables
like the symmetric BCH series
$\log(\ee^{\frac{1}{2}\AA}\ee^\BB\ee^{\frac{1}{2}\AA})$,
the BCH series with 3 generators
$\log(\ee^\AA\ee^\BB\ee^\CC)$,
or the more complex expression 
$\log(\ee^{\frac{1}{6}\BB}\ee^{\frac{1}{2}\AA}
\ee^{\frac{2}{3}\BB+\frac{1}{72}[B,[A,B]]}\ee^{\frac{1}{2}\AA}\ee^{\frac{1}{6}\BB})$
with a commutator in an exponential.


\section{Implementation}\label{Sec:Imnplementation}
Here we only describe the main ideas on which the implementation of \verb|bch| is based, details will be available in \cite{HHfast}.

The program \verb|bch| is implemented in the  C programming language according to the
C99 (or later) standard. The C code for parsing expressions of the form (\ref{eq:expression}) has
been generated by GNU Bison \cite{bison}.
The program uses  \verb|khash.h| from the \verb|klib| library \cite{klib}, 
which is a very efficient implementation of hash tables.
Apart from \verb|khash.h| it is self-contained, i.e., independent of external libraries except the standard C library. 

In particular, the program does not require a library for multi-precision integer or rational arithmetic.
Instead all calculations are carried out in  pure integer arithmetic using the 
128 bit integer type \verb|__int128_t|, which is available for most current C compilers on 
current hardware. With  this integer type computations of the BCH series up to degree
$N=30$ are possible.
To avoid calculations with rational numbers,
the program first determines a common denominator for all rational numbers that can 
occur during the computation (see \cite{HHdenom}, \cite{HHSmallestDenom}), 
then all calculations are reorganized in such a way
that they can actually be carried out in pure integer arithmetic.

In a  preliminary step,
the program  computes the coefficients of all Lyndon words up to degree 
$N$ 
 in the power series of the given expression $X$ of the form (\ref{eq:expression})
using an adaptation of the algorithm from \cite{HHetalWordAlg}. In the special
case $X=\log(\ee^\AA\ee^\BB)$ it alternatively uses the algorithm from 
the appendix of \cite{HHSmallestDenom} and exploits the symmetry that
the coefficient of the word $\AA^{q_1}\BB^{q_2}\cdots(\AA\lor\BB)^{q_m}$ is
invariant under permutations of the exponents $q_1,\dots,q_m$, see \cite{G}.

Next, the program transforms the coefficients of the Lyndon words into coefficients
of Lie basis elements in the Lie series.
For the Lyndon basis and the right-normed Chibrikov basis this is done by solving a linear system whose matrix consists of coefficients of Lyndon words in basis elements.

In the Lyndon basis case this matrix with integer entries is triangular with all diagonal
elements equal 1, which is a consequence of \cite[Theorem~5.1]{Reutenauer}. Thus, this
integer system is readily solved in integer arithmetic with no further denominators introduced.
Here the performance bottleneck is actually the computation of the coefficient matrix,
for which a new efficient algorithm has been developed, for details see \cite{HHfast}. 

In the right-normed Chibrikov basis case, the coefficient matrix is no longer
triangular, but it still has determinant $\pm 1$, and its leading principal
minors are all equal $\pm 1$. It follows that the matrix has a $LU$-decomposition
with triangular matrices $L$, $U$ whose entries are all integers and whose diagonal
elements are equal $\pm 1$. This $LU$-decomposition can be computed 
in pure integer arithmetic using Gaussian elimination without any row or column permutations.
Again solving such a linear system  introduces no further denominators.

In the Hall basis case, the program first computes the representation of
the Lie series in the Lyndon basis, and then uses a standard rewriting algorithm
(see \cite[Section~4.2]{Reutenauer})
applied to the Lyndon basis elements to obtain a representation in the Hall basis.
Because this rewriting algorithm is carried out in integer arithmetic, again no further
denominators are introduced.

The special case $X=\log(\ee^\AA\ee^\BB)$ allows a significant optimization in 
the Lyndon basis or right-normed Chibrikov basis case, if $N$ is even: 
Here the Lie series is computed using the above method only up to terms
of degree $N-1$. The terms of degree $N$ can then efficiently be computed using a
formula due to E. Eriksen, see \cite[Section~III.A]{Eriksen}.

As a final remark on the implementation we note that the computation can
be organized in such a way that finely homogeneous components 
$H_{(a,b)}$ of $H$ are computed completely independently of each other.
Here a finely homogeneous component $H_{(a,b)}$ of multi-degree $(a,b)$
is of degree $a$ in $\AA$ and of degree $b$ in $\BB$ such that
$$H_n = \sum_{a+b=n}H_{(a,b)}.$$
Thus, the computations of these components can be done in parallel, giving a
simple but effective parallelization strategy, which 
has been implemented using OpenMP.

\section{Installation}
The source code  can be downloaded from the GitHub repository
\begin{quote}
{\tt https://github.com/HaraldHofstaetter/BCH}
\end{quote}
On a Unix-like system with \verb|gcc| compiler available just type
\begin{quote}
\begin{BVerbatim}
$ make
\end{BVerbatim} 
\end{quote}
in a directory containing the source code, which causes 
the shared library \verb|libbch.so| and the executable \verb|bch|
to be created.
To use a different compiler, the \verb|Makefile| has to be adapted accordingly.

The \verb|bch| program has been compiled to WebAssembly~\cite{wasm}. 
So one can try it in the browser without installing it. 
For example, to compute the BCH series represented 
in the Lyndon basis up to terms of degree $N=20$ just 
open the following link:
\begin{center}
\verb|http://www.harald-hofstaetter.at/BCH/bch.html?verbosity_level=1&N=20&basis=0|
\end{center}
Parameters are specified as shown. 
Their usage is (up to obvious modifications, e.g.~`\verb|&|' instead of
whitespace) exactly as described in Section~\ref{Sec:usage_bch}.

\section{Usage of the {\tt bch} program}\label{Sec:usage_bch}
The program invoked without any arguments produces the following output:

\medskip

{\small\begin{BVerbatim}
$ ./bch
+1/1*A+1/1*B+1/2*[A,B]+1/12*[A,[A,B]]+1/12*[[A,B],B]+1/24*[A,[[A,B],B]]-1/720*[A
,[A,[A,[A,B]]]]+1/180*[A,[A,[[A,B],B]]]+1/360*[[A,[A,B]],[A,B]]+1/180*[A,[[[A,B]
,B],B]]+1/120*[[A,B],[[A,B],B]]-1/720*[[[[A,B],B],B],B]
\end{BVerbatim}
}

\medskip

\noindent This is the BCH series (i.e., the Lie series for $\log(\ee^{\AA}\ee^{\BB})$)
up to terms of degree $N=5$ represented in the Lyndon basis.

Arguments have the form
\begin{center}
{\tt\em parameter=value}
\end{center}
Such arguments may be combined arbitrarily. The following parameters are available 
(with default values in square brackets):

\subsection*{{\tt N[=5]}}
The maximal degree up to which the Lie series is computed.

\medskip

{\small\begin{BVerbatim}
$ ./bch N=8
+1/1*A+1/1*B+1/2*[A,B]+1/12*[A,[A,B]]+1/12*[[A,B],B]+1/24*[A,[[A,B],B]]-1/720*[A
,[A,[A,[A,B]]]]+1/180*[A,[A,[[A,B],B]]]+1/360*[[A,[A,B]],[A,B]]+1/180*[A,[[[A,B]
,B],B]]+1/120*[[A,B],[[A,B],B]]-1/720*[[[[A,B],B],B],B]-1/1440*[A,[A,[A,[[A,B],B
]]]]+1/720*[A,[[A,[A,B]],[A,B]]]+1/360*[A,[A,[[[A,B],B],B]]]+1/240*[A,[[A,B],[[A
,B],B]]]-1/1440*[A,[[[[A,B],B],B],B]]+1/30240*[A,[A,[A,[A,[A,[A,B]]]]]]-1/5040*[
A,[A,[A,[A,[[A,B],B]]]]]+1/10080*[A,[A,[[A,[A,B]],[A,B]]]]+1/3780*[A,[A,[A,[[[A,
B],B],B]]]]+1/10080*[[A,[A,[A,B]]],[A,[A,B]]]+1/1680*[A,[A,[[A,B],[[A,B],B]]]]+1
/1260*[A,[[A,[[A,B],B]],[A,B]]]+1/3780*[A,[A,[[[[A,B],B],B],B]]]+1/2016*[[A,[A,B
]],[A,[[A,B],B]]]-1/5040*[[[A,[A,B]],[A,B]],[A,B]]+13/15120*[A,[[A,B],[[[A,B],B]
,B]]]+1/10080*[[A,[[A,B],B]],[[A,B],B]]-1/1512*[[A,[[[A,B],B],B]],[A,B]]-1/5040*
[A,[[[[[A,B],B],B],B],B]]+1/1260*[[A,B],[[A,B],[[A,B],B]]]-1/2016*[[A,B],[[[[A,B
],B],B],B]]-1/5040*[[[A,B],B],[[[A,B],B],B]]+1/30240*[[[[[[A,B],B],B],B],B],B]+1
/60480*[A,[A,[A,[A,[A,[[A,B],B]]]]]]-1/15120*[A,[A,[A,[[A,[A,B]],[A,B]]]]]-1/100
80*[A,[A,[A,[A,[[[A,B],B],B]]]]]+1/20160*[A,[[A,[A,[A,B]]],[A,[A,B]]]]-1/20160*[
A,[A,[A,[[A,B],[[A,B],B]]]]]+1/2520*[A,[A,[[A,[[A,B],B]],[A,B]]]]+23/120960*[A,[
A,[A,[[[[A,B],B],B],B]]]]+1/4032*[A,[[A,[A,B]],[A,[[A,B],B]]]]-1/10080*[A,[[[A,[
A,B]],[A,B]],[A,B]]]+13/30240*[A,[A,[[A,B],[[[A,B],B],B]]]]+1/20160*[A,[[A,[[A,B
],B]],[[A,B],B]]]-1/3024*[A,[[A,[[[A,B],B],B]],[A,B]]]-1/10080*[A,[A,[[[[[A,B],B
],B],B],B]]]+1/2520*[A,[[A,B],[[A,B],[[A,B],B]]]]-1/4032*[A,[[A,B],[[[[A,B],B],B
],B]]]-1/10080*[A,[[[A,B],B],[[[A,B],B],B]]]+1/60480*[A,[[[[[[A,B],B],B],B],B],B
]]
\end{BVerbatim}
}

\subsection*{\tt basis[=0]}
With default value \verb|basis=0| the result is represented in the 
Lyndon basis;   with  value \verb|basis=1| in the right-normed Chibrikov basis:

\medskip

{\small\begin{BVerbatim}
$ ./bch basis=1
+1/1*A+1/1*B-1/2*[B,A]-1/12*[A,[B,A]]+1/12*[B,[B,A]]+1/24*[B,[A,[B,A]]]+1/720*[A
,[A,[A,[B,A]]]]-1/360*[B,[A,[A,[B,A]]]]+1/120*[A,[B,[A,[B,A]]]]-1/120*[B,[B,[A,[
B,A]]]]+1/360*[A,[B,[B,[B,A]]]]-1/720*[B,[B,[B,[B,A]]]]
\end{BVerbatim}
}

\medskip

\noindent With value \verb|basis=2| the result is represented in the classical Hall basis:

\medskip

{\small\begin{BVerbatim}
$ ./bch basis=2
+1/1*A+1/1*B-1/2*[B,A]+1/12*[[B,A],A]-1/12*[[B,A],B]+1/24*[[[B,A],A],B]-1/720*[[
[[B,A],A],A],A]-1/180*[[[[B,A],A],A],B]+1/180*[[[[B,A],A],B],B]+1/720*[[[[B,A],B
],B],B]-1/120*[[[B,A],A],[B,A]]-1/360*[[[B,A],B],[B,A]]
\end{BVerbatim}
}

\medskip


\subsection*{\tt generators[=\tt ABC\dots]}
This parameter enables alternative names for the generators:
\medskip

{\small\begin{BVerbatim}
$ ./bch generators=xy
+1/1*x+1/1*y+1/2*[x,y]+1/12*[x,[x,y]]+1/12*[[x,y],y]+1/24*[x,[[x,y],y]]-1/720*[x
,[x,[x,[x,y]]]]+1/180*[x,[x,[[x,y],y]]]+1/360*[[x,[x,y]],[x,y]]+1/180*[x,[[[x,y]
,y],y]]+1/120*[[x,y],[[x,y],y]]-1/720*[[[[x,y],y],y],y]
\end{BVerbatim}
}

\medskip

\subsection*{\tt expression[=0]}
The program can compute the Lie series of arbitrary Lie expressions of the 
form (\ref{eq:expression}). Some expressions are predefined:
\begin{itemize}[leftmargin=*]
\item {\tt expression=0} -- compute Lie series for $\log(\ee^\AA\ee^\BB)$, the classical BCH formula.
  \item {\tt expression=1} -- compute Lie series for $\log(\ee^{\frac{1}{2}\AA}\ee^\BB\ee^{\frac{1}{2}\AA})$, the symmetric BCH formula:

{\small\begin{BVerbatim}
$ ./bch expression=1
+1/1*A+1/1*B-1/24*[A,[A,B]]+1/12*[[A,B],B]+7/5760*[A,[A,[A,[A,B]]]]-7/1440*[A,[A
,[[A,B],B]]]+1/360*[[A,[A,B]],[A,B]]+1/180*[A,[[[A,B],B],B]]+1/120*[[A,B],[[A,B]
,B]]-1/720*[[[[A,B],B],B],B]
\end{BVerbatim}
}

\item {\tt expression=2} -- compute Lie series for $\log(\ee^\AA\ee^\BB\ee^\AA)$, another symmetric version of the BCH formula:

{\small\begin{BVerbatim}
$ ./bch expression=2
+2/1*A+1/1*B-1/6*[A,[A,B]]+1/6*[[A,B],B]+7/360*[A,[A,[A,[A,B]]]]-7/180*[A,[A,[[A
,B],B]]]+1/45*[[A,[A,B]],[A,B]]+1/45*[A,[[[A,B],B],B]]+1/30*[[A,B],[[A,B],B]]-1/
360*[[[[A,B],B],B],B]
\end{BVerbatim}
}

\item {\tt expression=3} -- compute Lie series for $\log(\ee^\AA\ee^\BB\ee^\CC)$, a BCH
formula with three different exponentials: 

{\small\begin{BVerbatim}
$ ./bch expression=3 N=4
+1/1*A+1/1*B+1/1*C+1/2*[A,B]+1/2*[A,C]+1/2*[B,C]+1/12*[A,[A,B]]+1/12*[A,[A,C]]+1
/12*[[A,B],B]+1/3*[A,[B,C]]+1/6*[[A,C],B]+1/12*[[A,C],C]+1/12*[B,[B,C]]+1/12*[[B
,C],C]+1/24*[A,[[A,B],B]]+1/12*[A,[A,[B,C]]]+1/12*[A,[[A,C],B]]+1/24*[A,[[A,C],C
]]+1/12*[A,[B,[B,C]]]+1/12*[[A,[B,C]],B]+1/12*[A,[[B,C],C]]+1/12*[[A,C],[B,C]]+1
/24*[B,[[B,C],C]]
\end{BVerbatim}
}

\item {\tt expression=4} -- compute Lie series for $\log(\ee^\AA\ee^\BB\ee^{-\AA}\ee^{-\BB})$:
    
{\small\begin{BVerbatim}
$ ./bch expression=4 
+1/1*[A,B]+1/2*[A,[A,B]]-1/2*[[A,B],B]+1/6*[A,[A,[A,B]]]-1/4*[A,[[A,B],B]]+1/6*[
[[A,B],B],B]+1/24*[A,[A,[A,[A,B]]]]-1/12*[A,[A,[[A,B],B]]]+1/12*[[A,[A,B]],[A,B]
]+1/12*[A,[[[A,B],B],B]]-1/24*[[[[A,B],B],B],B]
\end{BVerbatim}
}

\item {\tt expression=5} -- compute Lie series for $\log(\ee^{\frac{1}{6}\BB}\ee^{\frac{1}{2}\AA}
\ee^{\frac{2}{3}\BB+\frac{1}{72}[B,[A,B]]}\ee^{\frac{1}{2}\AA}\ee^{\frac{1}{6}\BB})$,
a more complicated version of the BCH formula, which is constructed in such a way that 
besides $\AA+\BB$ all terms have degree $\geq 5$:

{\small\begin{BVerbatim}
$ ./bch expression=5
+1/1*A+1/1*B+1/2880*[A,[A,[A,[A,B]]]]-7/8640*[A,[A,[[A,B],B]]]+1/2160*[[A,[A,B]]
,[A,B]]+7/12960*[A,[[[A,B],B],B]]+1/4320*[[A,B],[[A,B],B]]-41/155520*[[[[A,B],B]
,B],B]
\end{BVerbatim}
}
\end{itemize}
An arbitrary expression of the form (\ref{eq:expression}) can directly be specified
by its formula as 

\begin{center}
{\tt "expression={\em formula}"}
\end{center}

\noindent Here the double quotes are necessary because the formula usally contains
special characters like `\verb|*|' or brackets that confuse the command line shell.
For example the above Lie series for  $\log(\ee^{\frac{1}{6}\BB}\ee^{\frac{1}{2}\AA}
\ee^{\frac{2}{3}\BB+\frac{1}{72}[B,[A,B]]}\ee^{\frac{1}{2}\AA}\ee^{\frac{1}{6}\BB})$
can also be obtained in the following way:

\medskip

{\small\begin{BVerbatim}
$ ./bch "expression=log(exp(1/6*B)*exp(1/2*A)*exp(2/3*B+1/72*[B,[A,B]])*exp(1/2*
A)*exp(1/6*B))" 
+1/1*A+1/1*B+1/2880*[A,[A,[A,[A,B]]]]-7/8640*[A,[A,[[A,B],B]]]+1/2160*[[A,[A,B]]
,[A,B]]+7/12960*[A,[[[A,B],B],B]]+1/4320*[[A,B],[[A,B],B]]-41/155520*[[[[A,B],B]
,B],B]
\end{BVerbatim}
}

\medskip

\subsection*{\tt table\_output[=0 {\em or} =1 {\em depending on size of result}]}
With \verb|table_output=0| the result is displayed as a linear combination of commutators. 
This is the default if the result consists of less than 200 terms.
With
\verb|table_output=1|, on the other hand, the result is displayed in tabular form.
Each row of the table corresponds to a basis element of the respective Lie basis,
where by default the index, the indices of the left and
the right factors, and the coefficient of the basis element is displayed. 
This information uniquely determines the basis element, which follows from the
fact that 
for all the bases we consider (i.e., Lyndon, Hall, and right-normed Chibrikov) the factors $h', h''$ of basis elements $h=[h',h'']$ of degree $\geq 2$ are
basis elements themselves.
\begin{quote} %
{\small\begin{BVerbatim}
$ ./bch table_output=1
0       1       0       0       1/1
1       1       1       0       1/1
2       2       0       1       1/2
3       3       0       2       1/12
4       3       2       1       1/12
5       4       0       3       0/1
6       4       0       4       1/24
7       4       4       1       0/1
8       5       0       5       -1/720
9       5       0       6       1/180
10      5       3       2       1/360
11      5       0       7       1/180
12      5       2       4       1/120
13      5       7       1       -1/720
\end{BVerbatim}
}\end{quote}
The displayed items can be customized by specifying the following parameters
which should be self-explanatory:
\begin{itemize}
\item {\tt print\_index[=1]}
\item {\tt print\_degree[=1]}
\item {\tt print\_multi\_degree[=0]}
\item {\tt print\_factors[=1]}
\item {\tt print\_foliage[=0]}
\item {\tt print\_basis\_element[=0]}
\item {\tt print\_coefficient[=1]}
\end{itemize}
\begin{quote} %
{\small\begin{BVerbatim}
$ ./bch table_output=1 print_basis_element=1 print_multi_degree=1
0       1       (1,0)   0       0       A       1/1
1       1       (0,1)   1       0       B       1/1
2       2       (1,1)   0       1       [A,B]   1/2
3       3       (2,1)   0       2       [A,[A,B]]       1/12
4       3       (1,2)   2       1       [[A,B],B]       1/12
5       4       (3,1)   0       3       [A,[A,[A,B]]]   0/1
6       4       (2,2)   0       4       [A,[[A,B],B]]   1/24
7       4       (1,3)   4       1       [[[A,B],B],B]   0/1     
8       5       (4,1)   0       5       [A,[A,[A,[A,B]]]]       -1/720
9       5       (3,2)   0       6       [A,[A,[[A,B],B]]]       1/180
10      5       (3,2)   3       2       [[A,[A,B]],[A,B]]       1/360
11      5       (2,3)   0       7       [A,[[[A,B],B],B]]       1/180
12      5       (2,3)   2       4       [[A,B],[[A,B],B]]       1/120
13      5       (1,4)   7       1       [[[[A,B],B],B],B]       -1/720
\end{BVerbatim}
}\end{quote}
Note that if the parameter  \verb|verbosity_level| 
is set to a value $\geq 1$ (see below), then  a header line for the table 
is printed.

\subsection*{\tt verbosity\_level[=0]}
With \verb|verbosity_level| set to a value $\geq 1$ some useful information about
the performance of the computation is printed, for example the common denominator 
and timings of the main steps of the computation as described in Section~\ref{Sec:Imnplementation}. Furthermore a table showing the number of the
Lie basis elements and the number of nonzero coefficients of the Lie series is printed.
For easy identification, each line of this output has the character \verb|#| at the leading position.

With \verb|verbosity_level| set to a value $\geq 2$ this  output is not buffered but flushed immediately. 
This is only recommended if long running times of the program are expected.

\begin{quote} %
{\small\begin{verbatim}
$ ./bch N=20 verbosity_level=1 | head -n 60
#number of Lyndon words of length<=20 over set of 2 letters: 111013
#initialize Lyndon words: time=0.0344573 sec
#expression=log(exp(A)*exp(B))
#denominator=102181884343418880000
#compute Goldberg coefficients: time=0.0137833 sec
#compute coefficients of Lyndon words: time=0.0226991 sec
#convert to Lie series: time=0.158603 sec
#compute terms of degree 20: time=0.000395872 sec
#total time=0.216212 sec
# degree         dim    #nonzero   dim(cum.)   #nz(cum.)
#      1           2           2           2           2
#      2           1           1           3           3
#      3           2           2           5           5
#      4           3           1           8           6
#      5           6           6          14          12
#      6           9           5          23          17
#      7          18          18          41          35
#      8          30          17          71          52
#      9          56          55         127         107
#     10          99          55         226         162
#     11         186         186         412         348
#     12         335         185         747         533
#     13         630         630        1377        1163
#     14        1161         629        2538        1792
#     15        2182        2181        4720        3973
#     16        4080        2181        8800        6154
#     17        7710        7710       16510       13864
#     18       14532        7709       31042       21573
#     19       27594       27594       58636       49167
#     20       52377       27593      111013       76760
#
# multi-degree  dim	    #nonzero
# ( 1,19)       1       0
# ( 2,18)	      9	      1
# ( 3,17)	      57      9
# ( 4,16)	      240     51
# ( 5,15)	      775	    204
# ( 6,14)	      1932    612
# ( 7,13)	      3876    1428
# ( 8,12)	      6288    2652
# ( 9,11)	      8398    3978
# (10,10)	      9225    4862
# (11, 9)	      8398    4862
# (12, 8)	      6288    3978
# (13, 7)	      3876    2652
# (14, 6)	      1932    1428
# (15, 5)	      775     612
# (16, 4)	      240     204
# (17, 3)	      57      51
# (18, 2)	      9       9
# (19, 1)	      1       0
#
# i     |i|     i'      i"      coefficient
0       1       0       0       1/1
1       1       1       0       1/1
2       2       0       1       1/2
3       3       0       2       1/12
4       3       2       1       1/12
5       4       0       3       0/1
6       4       0       4       1/24
\end{verbatim}
}\end{quote}

\section{Performance of the {\tt bch} program}\label{Sect:performance}
We document the performance of typical runs of the \verb|bch|
program on an ordinary personal computer.
More specifically, the computer system was a
a 3.0\,GHz Intel Core i5-2320 processor with 4 CPU cores and
8\,GB of RAM running an Ubuntu Linux operating system. The 
C compiler was \verb|gcc| version 9.3.0.
The memory usage was measured with \verb|/usr/bin/time|.
For the test runs of the program we used different 
maximum degrees $N$ and different bases. The results
are as follows: (Here, dimension is the number of basis elements of degree $\leq N$
and \#nonzero is the number of coefficients $\neq 0$.)

\subsubsection*{Lyndon basis:}

\nopagebreak

\begin{center}
\begin{tabular}{rrrrr}
\hline
$N$ & dimension & \#nonzero & time  & memory \\
\hline
18 &  31042 & 21573 &  $0.04$\,sec & 4628\,kB\\
20 & 111013 & 76760 & $0.22$\,sec  & 10932\,kB \\
22 & 401428 & 276474 & $2.13$\,sec & 33008\,kB \\
24 & 1465020 & 1005917 & $54.43$\,sec  & 115\,MB\\
26 & 5387991 & 3690268 & 16\,min 16\,sec  &  427\,MB\\
28 & 19945394 & 13632278 & 3\,h 57\,min & 1.4\,GB\\
30 & 74248451 & 50657857 & 54\,h 46\,min & 5.5\,GB\\
\hline
\end{tabular}
\end{center}

\subsubsection*{Classical Hall basis:}
\begin{center}
\begin{tabular}{rrrrr}
\hline
$N$ & dimension & \#nonzero & time  & memory \\
\hline
18 & 31042 & 30477   & $0.35$\,sec & 81248\,kB\\
20 & 111013 & 109697 & $2.12$\,sec & 428\,MB \\
22 & 401428 & 398313 & 20\,min 53\,sec & 2.6\,GB\\
\hline
\end{tabular}
\end{center}

\subsubsection*{Right-normed Chibrikov basis:}
\begin{center}
\begin{tabular}{rrrrr}
\hline
$N$ & dimension & \#nonzero & time  & memory \\
\hline
18 & 31042  & 21561 & $0.76$\,sec & 58300\,kB \\
20 & 111013 & 76748 & $21.52$\,sec & 637\,MB \\
22 & 401428 & 276463 & 21\,min 53\,sec & 6.9\,GB\\
\hline
\end{tabular}
\end{center}

For the Lyndon basis, the maximum degree $N=30$ that the program
can in principle handle could actually be achieved.
For the other bases, $N=22$ was the limit due to memory constraints
(the computer system had 8\,GB of RAM).

It is natural to compare the performance of our program with that
of other implementations of algorithms for computing the BCH series
up to terms of high degree.
Ref.~\cite{CasasMurua} is the only recently published report 
of such an implementation
that we know of. 
Results obtained with this implementation are available at 
\begin{quote}
\verb|http://www.ehu.eus/ccwmuura/bch.html|.
\end{quote}    
It should be acknowledged that these results served us well in verifying the computations of our \verb|bch| program.

For the implementation of \cite{CasasMurua} in {\sc Mathematica}
a completely different and  more sophisticated 
approach than ours was used. This approach, which is
based on a Lie algebraic structure of labeled rooted trees,
has the  obvious disadvantage  that
it is not easily adapted to arbitrary expressions 
of the form (\ref{eq:expression}). Thus, only the classical and the symmetric BCH series are handled in \cite{CasasMurua}.

In \cite{CasasMurua} it is reported  that the computation of the BCH series
up to terms of degree $N=20$ in the classical Hall basis required  less than
15\,min CPU time and 1.5\,GB of memory 
on a 2.4\,GHz Intel Core 2 Duo processor with 2\,GB of RAM.
For the computation in the Lyndon basis it required 3.6\,GB of memory, but no
specific information about the CPU time is given, only that
it required more time than the computation in the classical Hall basis.
This has to be compared with  our implementation, which  requires
only 0.22\,sec CPU time and 11\,MB of RAM for the Lyndon basis, 
and 2.12\,sec CPU time and 428\,MB of RAM for the classical Hall basis.
These numbers speak for themselves, even if  direct comparisons 
of different implementations in different programming languages that run on different hardware 
must be interpreted with caution.

\section{Usage of the {\tt libbch} library}
The \verb|main()| function 
of  the executable \verb|bch|, which is
defined in the source file \verb|bch.c|, essentially does nothing else
than processing the command line arguments and then calling 
the appropriate functions from the \verb|libbch| library, which
perform the actual computations.
In the same way, these functions can also be called from a user-defined program,
which has to be linked to the 
shared library \verb|libbch.so| for this purpose.
Each such program should include the header file
\verb|bch.h|, which contains declarations 
for the functions and data structures provided by the library.
The source file \verb|bch.c| for the executable \verb|bch| can serve as a model for
the usage of the  \verb|libbch| library.

In the following we give an 
excerpt from the header file \verb|bch.h|
covering  essential declarations together with some
necessary explanations. 
It has been tried to use names for the functions and their arguments
that are as self-explanatory as possible, and that are consistent with
the names of the corresponding parameters for the  \verb|bch| program
as described in Section~\ref{Sec:usage_bch}.

\subsection{128 bit integer type}
Essentially all computations of the \verb|libbch| library are carried out in pure integer arithmetic, a
large part of them in 128 bit integer arithmetic (the other part can be done
using integers with fewer bits). In \verb|bch.h| 
the 128 bit integer type  \verb|INTEGER| is defined
as an alias for the built-in type \verb|__int128_t|:
\begin{verbatim}
typedef __int128_t INTEGER; 
\end{verbatim}
Although such 128 bit integers are available for most modern
C compilers, the C 
standard library does not
provide functions for converting such integers to strings, or print them 
on standard output. The \verb|libbch| library provides such functions and also functions 
for converting and printing rational numbers, which are 
specified by their numerators and denominators of type \verb|INTEGER|.
Note that the rational numbers are reduced to lowest terms first.
\begin{verbatim}
int str_INTEGER(char *str, INTEGER x);
int str_RATIONAL(char *str, INTEGER num, INTEGER den);
void print_INTEGER(INTEGER x);
void print_RATIONAL(INTEGER num, INTEGER den);
\end{verbatim}
The functions \verb|str_INTEGER| and \verb|str_RATIONAL| write
to a character string \verb|str| which has to be long enough 
to contain the output including a terminating \verb|'\0'|.
They return the number of characters written (excluding the terminating \verb|'\0'|).
If \verb|str| is \verb|NULL|, then these functions return the number
of characters that would have been written in case of a proper output
string \verb|str|.

\subsection{Lie series}
The following two functions compute respectively the classical BCH series
(i.e., the Lie series for $\log(\ee^{\AA}\ee^{\BB})$) and
the symmetric BCH series (i.e., the Lie series for 
$\log(\ee^{\frac{1}{2}\AA}\ee^{\BB}\ee^{\frac{1}{2}\AA})$:
\begin{verbatim}
lie_series_t* BCH(int maximum_degree, int basis);
lie_series_t* symBCH(int maximum_degree, int basis);
\end{verbatim}
Here the argument \verb|maximum_degree| specifies the maximum degree
up to which terms of the Lie series shall be computed. The argument
\verb|basis| specifies the basis in which the resulting Lie series
shall be represented, where the values 0,1,2 respectively correspond to the
Lyndon basis, the right-normed Chibrikov basis, and the classical Hall basis.
These functions return a pointer to a structure of type \verb|lie_series_t| which 
contains the resulting Lie series. 
This structure should be considered as an opaque data type for 
which access to its data is provided by data access functions (see below).

\begin{verbatim}
void set_verbosity_level(int verbosity_level);
\end{verbatim}
If the verbosity level is set to a value $\geq 1$ {\em before} calling
the functions \verb|BCH| or \verb|symBCH|, then these function print
some useful information about the performance of the computation to the standard output.

\subsection{Data access functions}
\begin{verbatim}
int dimension(lie_series_t *LS);
int maximum_degree(lie_series_t *LS);
int number_of_generators(lie_series_t *LS);
INTEGER denominator(lie_series_t *LS);
INTEGER numerator_of_coefficient(lie_series_t *LS,  int i);
int degree(lie_series_t *LS, int i);
int degree_of_generator(lie_series_t *LS, int i, uint8_t g);
int left_factor(lie_series_t *LS, int i);
int right_factor(lie_series_t *LS, int i);
int str_foliage(char *str, lie_series_t *LS, int i, char *generators);
int str_basis_element(char *str, lie_series_t *LS, int i, char *generators);
int str_coefficient(char *str, lie_series_t *LS, int i);
void print_foliage(lie_series_t *LS, int i, char *generators);
void print_basis_element(lie_series_t *LS, int i, char *generators);
void print_coefficient(lie_series_t *LS,  int i);
\end{verbatim}
Here, the argument \verb|LS| is a pointer which was
previously returned by the functions \verb|BCH| or \verb|symBCH|
(or \verb|lie_series|, see below).
The argument \verb|i| is the index of the basis element, for which
the respective information is requested.
The argument \verb|g| in \verb|degree_of_generator| specifies
the (index of the) generator for which the degree in the basis element with
index \verb|i| is requested.
The argument \verb|generators| should be a string like \verb|"ABCDEFG"| containing at 
position 0 the name for generator 0, at position 1 the name for generator 1, etc.

The functions \verb|str_foliage|, \verb|str_basis_element|, and \verb|str_coefficient| write
to a character string \verb|str| which has to be long enough 
to contain the output including a terminating \verb|'\0'|.
They return the number of characters written (excluding the terminating \verb|'\0'|).
If \verb|str| is \verb|NULL|, then these functions return the number
of characters that would have been written in case of a proper output
string \verb|str|.

The functions \verb|left_factor| and \verb|right_factor|
respectively
return the index of the left factor $h'$ and the index of the right
factor $h''$ of the basis element $h=[h',h'']$ with index \verb|i|.

\subsection{Cleaning up}
\begin{verbatim}
void free_lie_series(lie_series_t *LS);
\end{verbatim}

\subsection{User-defined expressions}\label{SubSec:UserDefExpr}
Besides Lie series for  $\log(\ee^{\AA}\ee^{\BB})$ and
$\log(\ee^{\frac{1}{2}\AA}\ee^{\BB}\ee^{\frac{1}{2}\AA})$, also 
Lie series for arbitrary expressions of the form (\ref{eq:expression})
can be computed.
Such expressions are represented as  binary expression
trees, for the generation of which  the \verb|libbch| library
provides the following functions:
%
\begin{verbatim}
expr_t* identity(void);
expr_t* generator(uint8_t g);
expr_t* sum(expr_t* arg1, expr_t* arg2);
expr_t* difference(expr_t* arg1, expr_t* arg2);
expr_t* product(expr_t* arg1, expr_t* arg2);
expr_t* negation(expr_t* arg);
expr_t* term(int numerator, int denominator, expr_t* arg);
expr_t* exponential(expr_t* arg);
expr_t* logarithm(expr_t* arg);
expr_t* commutator(expr_t* arg1, expr_t* arg2);
\end{verbatim}
These functions return a pointer to a structure of type \verb|expr_t|
which represents a node of the expression tree and contains data,
depending on the type of the node,
like pointers to subexpressions, the index of a generator, or numerator and
denominator of a rational coefficient.
For example, the following code generates the 
expression 
$\log(\ee^{\frac{1}{6}\BB}\ee^{\frac{1}{2}\AA}
\ee^{\frac{2}{3}\BB+\frac{1}{72}[B,[A,B]]}\ee^{\frac{1}{2}\AA}\ee^{\frac{1}{6}\BB})$:
\begin{quote} %
{\small\begin{BVerbatim}
expr_t *A = generator(0);
expr_t *B = generator(1);
expr_t* expression =
    logarithm(product(product(product(product(
    exponential(term(1, 6, B)), exponential(term(1, 2, A))),
    exponential(sum(term(2, 3, B), 
                    term(1, 72, commutator(B, commutator(A, B)))))), 
    exponential(term(1, 2, A))), exponential(term(1, 6, B))));
\end{BVerbatim}
}
\end{quote}
Alternatively to the explicit construction of expression trees using the
above functions, such expressions can also be obtained by parsing
an input string for which the \verb|libbch| library
provides the following function:
\begin{verbatim}
expr_t* parse(char *input, char *generators, int *number_of_generators);
\end{verbatim}
Here, the argument \verb|input| is the input string and \verb|generators| is an array of
characters of length at least the number of different generators 
(noncommuting variables) that occur in \verb|input|. 
On output the array \verb|generators| contains the symbols for these generators
in alphabetical order, and the variable pointed to by \verb|number_of_generators| 
contains the number of these generators. For example, the following code likewise
generates the expression 
$\log(\ee^{\frac{1}{6}\BB}\ee^{\frac{1}{2}\AA}
\ee^{\frac{2}{3}\BB+\frac{1}{72}[B,[A,B]]}\ee^{\frac{1}{2}\AA}\ee^{\frac{1}{6}\BB})$:
\begin{quote} %
{\small\begin{BVerbatim}
char gens[2];
int num_gen;
expr_t* expression = parse("log(exp(1/6*B)*exp(1/2*A)
                            *exp(2/3*B+1/72*[B,[A,B]])
                            *exp(1/2*A)*exp(1/6*B))", gens, &num_gen);
\end{BVerbatim}
}
\end{quote}
After executing this code the array \verb|gens| contains the characters 
\verb|A| and \verb|B| at positions 0 and 1, respectively, and the variable
\verb|num_gen| has the value 2.

It is possible to construct more general expressions than those  
of the form (\ref{eq:expression}), which cannot necessarily be represented 
as a Lie series, so that the result of the function {\tt lie\_series} described below
would not necessarily make any sense. Therefore it is recommended to use the following 
function for
checking such expressions whether they can actually be represented as  Lie series:
\begin{verbatim}
int is_lie_element(expr_t* expression);
\end{verbatim}

Finally, for an expression obtained in one of the above ways, the following function
computes the corresponding Lie series:
%
%
%
\begin{verbatim}
lie_series_t* lie_series(int number_of_generators, expr_t* expression, 
                         int maximum_degree, int basis);
\end{verbatim}
Regarding arguments and return value, this function is analogous to the functions \verb|BCH| and \verb|symBCH|,
but with additional arguments \verb|number_of_generators| and \verb|expression|. 
Here, \verb|number_of_generators| is the number of generators occurring in the expression (or more
precisely the highest index plus 1 of all generators occurring in the expression). In the above examples
it should have the value 2.

For cleaning up, the following function frees all memory that 
was allocated during the construction of all expressions
since the program was started (or since the last time this function was called):
\begin{verbatim}
void free_all_expressions(void);
\end{verbatim}
Usually it is not necessary to call this function explicitly because 
the \verb|libbch| library registers it with the standard library function \verb|atexit|
so that it is  automatically called when the program ends.

\end{document}